# SECURED COLOR IMAGE WATERMARKING TECHNIQUE IN DWT-DCT DOMAIN

Baisa L. Gunjal[1] and Suresh N.Mali[2]

[1]Amrutvahini College of Engineering, Sangamner, A'nagar, MS, India
`hello_baisa@yahoo.com`
[2]Imperial College of Engineering and Research, Wagholi, Pune, MS, India
`snmali@rediffmail.com`

## ABSTRACT

*The multilayer secured DWT-DCT and YIQ color space based image watermarking technique with robustness and better correlation is presented here. The security levels are increased by using multiple pn sequences, Arnold scrambling, DWT domain, DCT domain and color space conversions. Peak signal to noise ratio and Normalized correlations are used as measurement metrics. The 512x512 sized color images with different histograms are used for testing and watermark of size 64x64 is embedded in HL region of DWT and 4x4 DCT is used. 'Haar' wavelet is used for decomposition and direct flexing factor is used. We got PSNR value is 63.9988 for flexing factor k=1 for Lena image and the maximum NC 0.9781 for flexing factor k=4 in Q color space. The comparative performance in Y, I and Q color space is presented. The technique is robust for different attacks like scaling, compression, rotation etc.*

## KEYWORDS

DCT-DWT, Scrambling, Histogram, YIQ color space.

## 1. INTRODUCTION

It has become a daily need to create copy, transmit and distribute digital data as a part of widespread multimedia technology in internet era. Hence copyright protection has become essential to avoid unauthorized replication problem. Digital image watermarking provides copyright protection to image by hiding appropriate information in original image to declare rightful ownership [1]. Robustness, Perceptual transparency, capacity and Blind watermarking are four essential factors to determine quality of watermarking scheme [4][5]. Watermarking algorithms are broadly categorized as Spatial Domain Watermarking and Transformed domain watermarking. In spatial domain, watermark is embedded by directly modifying pixel values of cover image. Least Significant Bit insertion is example of spatial domain watermarking. In Transform domain, watermark is inserted into transformed coefficients of image giving more information hiding capacity and more robustness against watermarking attacks because information can be spread out to entire image [1]. Watermarking using Discrete Wavelet Transform, Discrete Cosine Transform, CDMA based Spread Spectrum Watermarking are examples of Transform Domain Watermarking. The rest of the paper is organized as follows: Section 2 focuses on survey of color image watermarking algorithms. Section 3 gives details of fundamentals of elements used in proposed methodology. In section 4, proposed methodology is explained. Section 5 shows Experimental results after implementation and Testing. In section 6, conclusion is drawn.

## 2. SURVEY

Though Fourier transform, short time Fourier transform and continuous wavelet transform are available in transform domain, but all of them are having their own limitations. Discrete Wavelet Transform provides multi resolution for given image and can efficiently implemented using digital filters, it has become attraction of researchers in image processing area. Here, review of literature survey is done on different transform in transform domain and existing color image watermarking techniques with based on 'Discrete Wavelet Transform. Following are some existing methods for in color image watermarking:' In [10], Integer Wavelet Transform with Bit Plane complexity Segmentation is used with more data hiding capacity. This method used RGB color space for watermark embedding. In [2] DWT based watermarking algorithm of color images is proposed. The RGB color space is converted into YIQ color space and watermark is embedded in Y and Q components. This method gives correlation up to 0.91 in JPEG Compression attack. In [3], Watermarking Algorithm Based on Wavelet and Cosine Transform for Color Image is proposed. A binary image as watermark is embedded into green or blue component of color image. In [4], Color Image Watermarking algorithm based on DWT-SVD is proposed. The scrambling watermark is embedded into green component of color image based on DWT-SVD. The scheme is robust and giving PSNR up to 42, 82. In [5], Pyramid Wavelet Watermarking Technique for Digital Color Images is proposed. This algorithm gives better security and better correlation in Noise and compression attacks.

## 3. FOUNDATIONS OF OUR METHODOLOGY

### 3.1. RGB and YIQ Color Space

RGB color space can be converted into YIQ color space. Y' is similar to perceived luminance; 'I and Q' carry color information and some luminance information. Since pixel values are highly correlated in RGB color spaces, the watermark embedding in YIQ color space is preferred for Watermarking. Initially color image is read and R, G, B components of original Cover Image are separated. Then they are converted into YIQ color Space using following equations [2]. After conversion of RGB color spaces into YIQ color spaces, Watermark is embedded.

$$Y = 0.299 * R + 0.587 * G + 0.114 * B \quad (1)$$
$$I = 0.596 * R - 0.274 * G - 0.322 * B \quad (2)$$
$$Q = 0.211 * R - 0.522 * G + 0.311 * B \quad (3)$$

After embedding the watermark using DWT, YIQ color space is converted back into RGB color space using following equations.

$$R = Y + 0.956 * I + 0.621 * Q \quad (4)$$
$$G = Y - 0.272 * I - 0.647 * Q \quad (5)$$
$$B = Y - 1.106 * I + 1.702 * Q \quad (6)$$

### 3.2. Selection of sub band in DWT

ISO has developed and generalized still image compression standard JPEG2000 which substitutes DWT for DCT. DWT offers mutiresolution representation of image and DWT gives perfect reconstruction of decomposed image. Discrete wavelet can be represented as

$$\psi_{j,k}(t) = a_0^{-j/2} \psi(a_0^{-j} t - k b_0) \quad (7)$$

For dyadic wavelets $a_0 = 2$ and $b_0 = 1$, Hence we have,

$$\psi_{j,k}(t) = 2^{-j/2} \psi(2^{-j} t - k) \quad j, k \in Z \quad (8)$$

When image is passed through series of low pass and high pass filters, DWT decomposes the image into sub bands of different resolutions [6][7][8][9].

| LL1 | HL1 |
|-----|-----|
| LH1 | HH1 |

Figure 1: One Level Image Decomposition

As shown in figure 1, DWT decomposes image into four non overlapping multi resolution sub bands: LL1 (Approximate sub band), HL1 (Horizontal sub band), LH1 (Vertical sub band) and HH1 (Diagonal Sub band). Here, LL1 is low frequency component whereas HL1, LH1 and HH1 are high frequency (detail) components [2].To obtain next coarser scale of wavelet coefficients after level 1, the sub band LL1 is further decomposed as per requirement. Embedding watermark in low frequency coefficients can increase robustness significantly but maximum energy of most of the natural images is concentrated in approximate (LL1) sub band. Hence modification in this low frequency sub band will cause severe and unacceptable image degradation. Hence watermark is not embedded in LL1 sub band. The good areas for watermark embedding are high frequency sub bands (HL1, LH1 and HH1), because human naked eyes are not sensitive to these sub bands. They yield effective watermarking without being perceived by human eyes. But HH1 sub band includes edges and textures of the image. Hence HH1 is also excluded. The rest options are HL1 and LH1. But Human Visual System is less sensitive in horizontal than vertical. Hence Watermarking done in HL1 region.

### 3.3. Selection of Block size in Discrete Cosine Transform

The discrete cosine transform (DCT) represents an image as a sum of sinusoids of varying magnitudes and frequencies. The DCT has special property that most of the visually significant information of the image is concentrated in just a few coefficients of the DCT [3]. It's referred as 'Energy compaction Property'. The DCT for image A with M x N size is given by:

$$DCT_{pq} = \alpha_p \alpha_q \sum_{m=0}^{M-1} \sum_{n=0}^{N-1} A_{mn} \cos\left(\frac{\pi(2m+1)p}{2M}\right) \cos\left(\frac{\pi(2n+1)q}{2N}\right) \quad (9)$$

where,

$$0 \leq p \leq M-1, \text{ and } 0 \leq q \leq N-1 \quad (10)$$

$$\alpha_p = \begin{cases} 1/\sqrt{M}, & p = 0 \\ \sqrt{2/M}, & 1 \leq p \leq M-1 \end{cases} \quad (11)$$

$$\alpha_q = \begin{cases} 1/\sqrt{N}, & q = 0 \\ \sqrt{2/N}, & 1 \leq q \leq N-1 \end{cases} \quad (12)$$

As DCT is having good energy compaction property, many DCT based Digital image watermarking algorithms are developed. It's already proved that DWT-DCT combined approach can significantly improve PSNR with compared to only DCT based watermarking methods.

### 3.4. Scrambling Method and Arnold Periodicity

Different methods can be used for image scrambling such as Fass Curve, Gray Code, Arnold Transform, Magic square etc. Here Arnold Transform is used. The special property of Arnold Transform is that image comes to it's original state after certain number of iterations. These' number of iterations' are called 'Arnold Period' or 'Periodicity of Arnold Transform'. The Arnold Transform of image is given by

$$\begin{pmatrix} x_n \\ y_n \end{pmatrix} = \begin{bmatrix} 1 & 1 \\ 1 & 2 \end{bmatrix} \begin{pmatrix} x \\ y \end{pmatrix} (mod\ N) \qquad (13)$$

Where, (x, y) ={0,1,.....N} are pixel coordinates from original image. ($x_n, y_n$) are corresponding results after Arnold Transform. The periodicity of Arnold Transform (P), is dependent on size of given image. From equation: 3 we have,

$x_n$=x+y  (14)
$y_n$=x+2*y

If (mod ($x_n$, N) ==1 && mod ($y_n$, N) ==1)  (15)

then P=N  (16)

## 4. PROPOSED METHODOLOGY

The Watermark Embedding and Extraction Process for HL sub band for I component is given below. Same procedure is followed applied for Y and Q components for results.

### 4.1 Watermark Embedding Algorithm

Step 1: Read Color Cover Image of 512x512 size. Separate it's R,G,B components and convert into YIQ color space using equations1,2,3.

Step 2: Now select I component and apply one level DWT. Consider HL1 sub band.

Step 3. Read grey scale watermark of 64x64 size.

Step 4: Depending upon Key K1, generate pn sequence for given watermark and calculate sum say SUM, which is summation of all elements in generated pn sequence.

Step 5. Determine Arnold Periodicity P for given watermark.

Step 6: If SUM > T, where T is some predefined threshold value, then perform watermark scrambling by Key K2= P+ Count, Otherwise perform watermark scrambling by Key K3= P+ Count, where count is predefined value used as counter. Here, we get 'Scrambled Watermark' by Arnold Transform.

Step 7: Generate two pn sequences: pn_sequence_0 and pn_sequence_1, depending upon sum of all elements of mid band used for 4x4 DCT transformation.

Step 8: Perform watermark embedding using following equations:

If Watermark bit is 0, then
$D' = D + K * pn\_sequence\_0$  (18)

If Watermark bit is 1, then
$D' = D + K * pn\_sequence\_1$  (19)

Where D is matrix of mid band coefficients of DCT Transformed block and $D'$ is Watermarked DCT block.

Step 9: Apply Inverse DCT to get 'New_HL1' component.

Step 10: Apply inverse DWT with 'LL1, New_HL1,LH1, HH1' to get 'New_I' component.

Step 11: Combine, Y, New_I and Q components and convert to RGB color space using equation 4, 5,6.

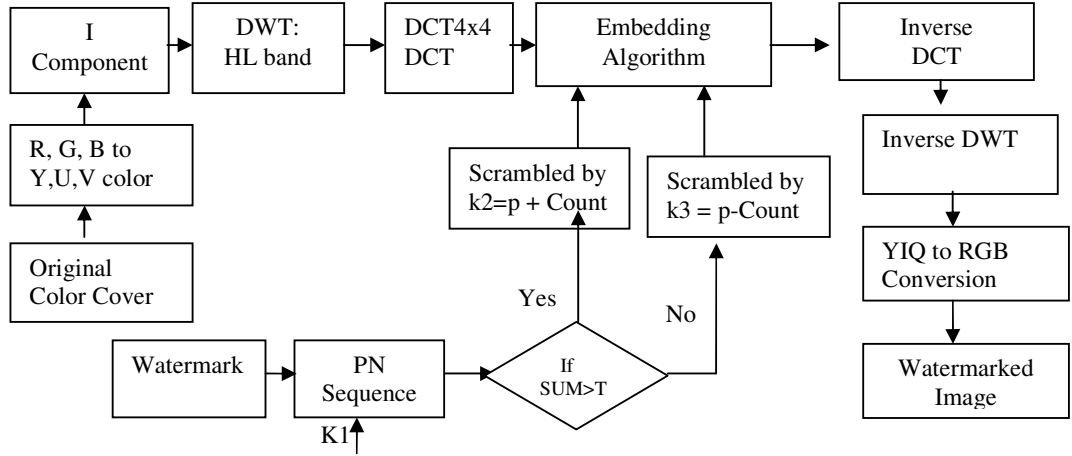

Figure 2: Block diagram for Watermark Embedding Process

## 4.2 Watermark Extraction Algorithm

Step 1: Read Color 'Watermarked_Image' and separate it's R,G and B components. Now convert to YIQ color space.

Step 2: Now select I component and apply one level DWT to retrieve HL1 sub band.

Step 3: Use 4x4 size for DCT blocks. Generate two pn sequences: pn_sequence_0 and pn_sequence_1, depending upon sum of all elements of mid band used for 4x4 DCT transformation. Use same seed which was used in watermark embedding process. e.g. if rand ('state', 15) is used in embedding process, then, same process is to repeated here.

Step 4: Extract mid band elements from DCT block and find correlation between 'extracted mid band coefficients and pn_sequence_0' as well as 'extracted mid band coefficients and pn_sequence_1'.

Step 5 Determine watermark bits as follows:
If correlation between 'extracted mid band coefficients and pn_sequence_0' is greater than 'extracted mid band coefficients and pn_sequence_1', then record watermark bit as 0 else record watermark bit as 1. Here we get 'Intermediate watermark'.

Step 6: Apply Arnold Scrambling to Intermediate watermark' to give final recovered watermark.

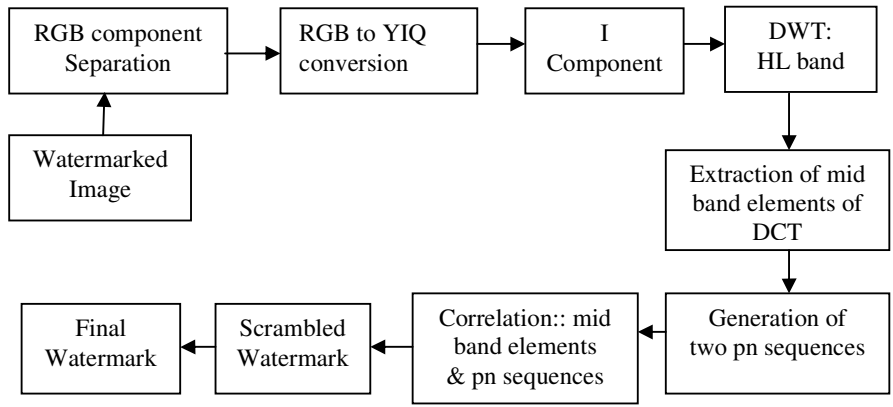

Figure 3: Block diagram for Watermark Extraction Process

## 5. EXPERIMENTATION

Image histograms are used as important feature during the testing of this method. Image histogram describes the distribution of pixel values into equal-sized bins. Then the range of pixels of image falling into each bin is calculated. The style of histogram may be described by:

$$H = \{h(i) | i = 1,2, \ldots 256|\} \tag{20}$$

where H is a vector denoting the volume-level histogram of intensity signal $F = \{f(i) | i = 1,2 \ldots N|\}$ and $h(i), h(i) \geq 0$ denotes number of samples in bin i and satisfies $\sum_{i=1}^{N} h(i) = N$
Different images have different pixel distribution. Hence they have different histogram shapes. The images having difference in their histogram shapes reflect better effect of algorithm. In figure 4, five different cover images of 512x512 sizes with their R, G, B histograms are shown. They are used as 'Test Cover Images' in our experiment. The watermark is of 64x64 size. The result is tested for flexing factor k=1, 2, 3,4. The sample results are presented only for k=1 and k=4 as shown in Table 1 and Table 2. For Lena image maximum PSNR value is 63.9988 for flexing factor k=1. The maximum NC 0.9781 for flexing factor k=4. The PSNR and NC values for Q channel are better than PSNR and NC values for Y and I channels. The experimentation is done in Matlab and PSNR (Peak Signal to Noise Ratio) and NC (Normalized Correlation) are used are measurement metrics. PSNR measures perceptual transparency' and given by:

$$PSNR(db) = 10 log_{10} \frac{(Max_I)^2}{\frac{1}{M*N}\sum_{i=1}^{M}\sum_{j=1}^{N}[f(i,j)-f'(i,j)]^2} \tag{21}$$

Where, f (i, j) is pixel of original image. f '(i, j) is pixel values of watermarked image. $Max_I$ is the maximum pixel value of image. NC measures robustness and given by:

$$NC = \frac{\sum_{i=1}^{N} w_i w_i{'}}{\sqrt{\sum_{i=1}^{N} w_i}\sqrt{\sum_{i=1}^{N} w_i{'}}} \tag{22}$$

where, N is total pixels in watermark, wi is original watermark, wi' is extracted watermark.

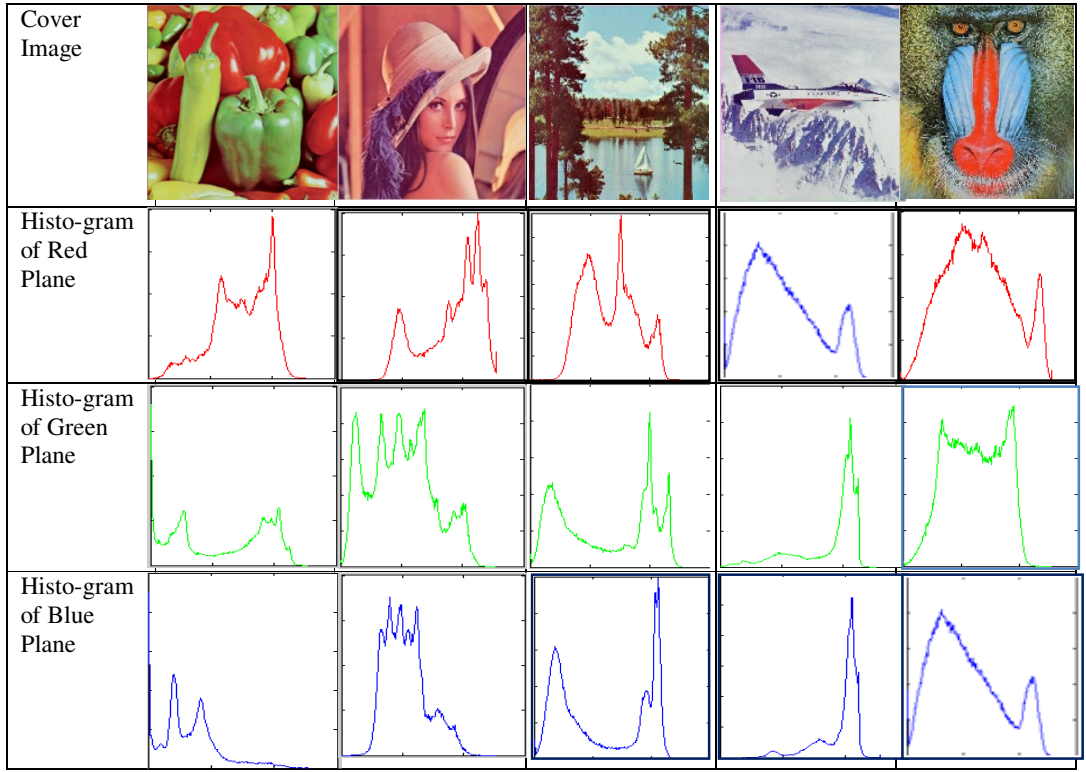

Figure 4: Images used as 'cover images' with histograms of Red Green and Blue Plane

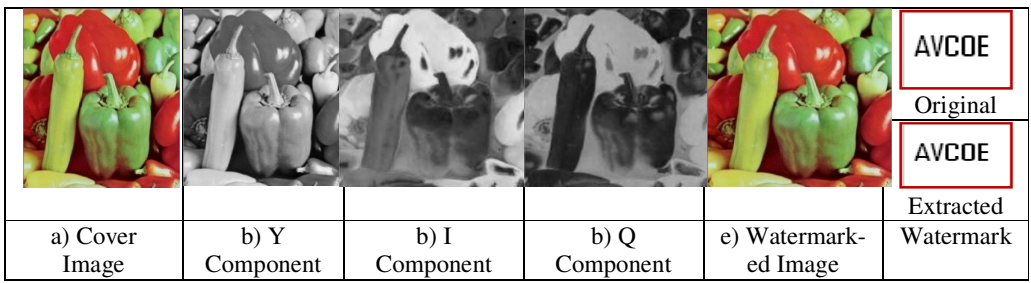

| a) Cover Image | b) Y Component | b) I Component | b) Q Component | e) Watermark-ed Image | Original / Extracted Watermark |

Figure 5: Cover Image 'peppers' and Y, I, Q components and Watermarked image, and watermark

| Cover Image | | | | | | | | | | |
|---|---|---|---|---|---|---|---|---|---|---|
| Metric | PSNR | NC | PSNR | NC | PSNR | NC | PSNR | NC | PSNR | NC |
| Y | 55.4919 | 0.3328 | 55.3712 | 0.3306 | 55.2730 | 0.2404 | 55.347 | 0.314 | 55.399 | 0.1278 |
| I | 55.9477 | 0.8111 | 55.8494 | 0.9266 | 55.6905 | 0.7077 | 55.450 | 0.903 | 55.816 | 0.7312 |
| Q | 63.1887 | 0.7425 | 63.9988 | 0.9355 | 64.6283 | 0.7342 | 67.568 | 0.916 | 63.947 | 0.7788 |

Table 1: Results of HL3 sub band for Flexing Factor k=1

| Cover Image | 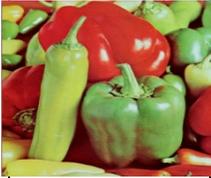 | | 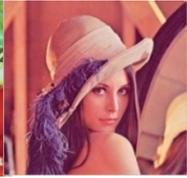 | | 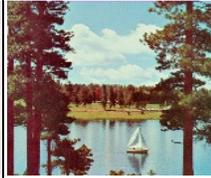 | | 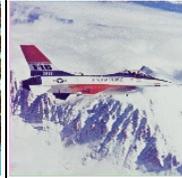 | | 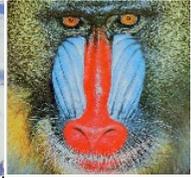 | |
|---|---|---|---|---|---|---|---|---|---|---|
| Metric | PSNR | NC | PSNR | NC | PSNR | NC | PSNR | NC | PSNR | NC |
| Y | 44.7861 | 0.7157 | 44.8162 | 0.7669 | 44.8008 | 0.5035 | 44.78 | 0.613 | 44.8000 | 0.3477 |
| I | 45.1319 | 0.9763 | 45.0086 | 0.9781 | 44.9074 | 0.9746 | 44.92 | 0.978 | 44.9621 | 0.9727 |
| Q | 48.6263 | 0.9781 | 48.5319 | 0.9781 | 48.4816 | 0.9781 | 48.46 | 0.978 | 48.5158 | 0.9781 |

Table 2: Results of HL3 sub band for Flexing Factor k=4

## 6. CONCLUSION

In this paper a strongly robust and multilayer security based color image watermarking algorithm in DWT-DCT domain is presented. Since pixel values are highly correlated in RGB color spaces, the use of YIQ color space for watermark embedding is beneficial for improvement in results. The images having difference in their histogram shapes reflect better effect of algorithm. Hence images with different histograms are used in experiment. The PSNR and NC values for Q channel are better than PSNR and NC values for Y and I channels. For Lena image the PSNR value is 63.9988 for flexing factor k=1. The maximum NC 0.9781 for flexing factor k=4. The average performance of PSNR and NC for Y and I channels are approximately remains in same range for given flexing factor. The technique is robust for different attacks like scaling, compression, rotation etc. This algorithm provides multilayer security by using pn sequence, Arnold scrambling.DWT domain, DCT domain, and color space conversions.

### ACKNOWLEDGMENT

Thanks to BCUD, Pune for providing 'Research Grant' for this work. File Ref. No.-BCUD/OSD/390 Dated 25/10/2010. We are thankful to 'Amrutvahini College of Engineering, Sangamner, A'nagar' and 'Imperial College of Engineering and Research', Wagholi, MS, India for providing technical support during the work.

### REFERENCES

[1] Cheng-qun Yin et al, " Color Image Watermarking Algorithm Based on DWT-SVD", Proceeding of the IEEE International Conference on Automation and Logistiocs August 18-21, 2007, Jinan, China, PP: 2607-2611

[2] Guangmin Sun, Yao Yu, " DWT Based Watermarking Algorithm of Color Images", Second IEEE Conference on Industrial Electronics and Application",2007, PP 1823-1826.

[3] Wei-Min Yang, Zheng Jin, " A Watermarking Algorithm Based on Wavelet and Cosine Transform for Color Image", First International Workshop on Education Technology and Computer Science", 2009, PP 899,903.

[4] Cheng-qun Yin, Li Li, An-qiang Lv and Li Qu, " Color Image Watermarking Algorithm Based on DWT-SVD", Proceeding of the IEEE International Conference on Automation and Logistics , August 18-21, 2007, Jinan, China, PP 2607-2611.

International Journal of Computer Science, Engineering and Information Technology (IJCSEIT), Vol.1, No.3, August 2011

[5] Awad Kh. Al-smari and Farhan A. Al-Enizi, "A Pyramid-Based Technique for Digital Color Images Copyright Protection", International Conference on Computing, Engineering and Information, 2009, pp 44-47

[6] B.L.Gunjal, R.R.Manthalkar, "Discrete Wavelet Transform Based Strongly Robust Watermarking Scheme for Information Hiding in Digital Images", Third International Conference- Emerging Trends in Engineering and Technology,19-21 Nov 2010 , Goa, India, ISBN 978-0-7695-4246-1, http://doi.ieeecomputersociety.org/10.1109/ICETET.2010.12.

[7] S. Joo, Y. Suh, J. Shin, H. Kikuchi, and S. J. Cho., "A new robust watermark embedding into wavelet DC components," ETRI Journal, 24, 2002, pp. 401-404.

[8] Voloshynovskiy. S. S. Pereira and T. Pun. 2001. "Attacks on Digital watermarks: classification, Estimation-Based attacks and Benchmarks", Comm, Magazine. 39(8):118-126.

[9] Abu-Errub, A., Al-Haj, A.,"Optimized DWT-based image watermarking", First International Conference on Applications of Digital Information and Web Technologies,  IEEE,2008, 4-6.

[10] K.Ramani, E Prasad, S Varadarajan, "Stenography using BPCS to the integer wavelet transform",IJCSNS international journal of Computer science and network security, vol-7,No: 7 July 2007.

## Authors

B.L.Gunjal completed her B.E. Computer from University of Pune and M.Tech in I.T. in Bharati  Vidyapeeth, Pune , Maharashtra, india. She is having 13 Years teaching experience and 18 international and national publications. Presently she is working on research project  on "Image Watermarking" funded by BCUD, University of Pune. Her areas of interest includes Image Processing, Advanced databases and Computer Networking.

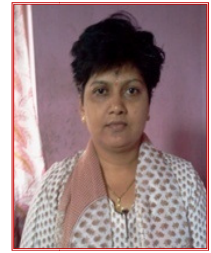

Dr. Suresh N. Mali has completed his PhD form Bharati Vidyapeeth Pune, presently he is working as Principal in Imperial College of Engineering and Research ,Wagholi, Pune. He is author of 3 books and having more than 18 international and national publications. He is working as Member of expert Committee of AICTE, New Delhi and also working as BOS member for Computer Engineering at University of Pune. He has worked as a Member of Local Inquiry Committee to visit at various institutes on behalf of University of Pune. His areas of interest mainly includes Image Processing

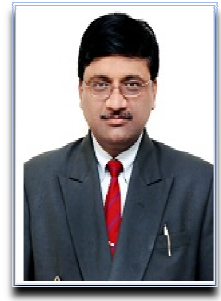